\documentclass[reprint, aps, prb, mathserif, superscriptaddress, showpacs, preprintnumbers]{revtex4-1}
\usepackage{times}
\usepackage{graphicx}
\usepackage{amsmath}
\usepackage{lineno}

\usepackage{natmove}
\usepackage [english]{babel}
\usepackage{float}
\usepackage [autostyle, english = american]{csquotes}
\MakeOuterQuote{"}

\newcommand{\eqn}[2]{\begin{equation}\label{#2}#1\end{equation}}
\newcommand{\mt}[1]{\mathrm{#1}}

\begin{document}



\title{Thermal boundary conductance across epitaxial ZnO/GaN interfaces: Assessment of phonon gas models and atomistic Green's function approaches for predicting interfacial phonon transport}

\author{John T. Gaskins}
\affiliation{Department of Mechanical and Aerospace Engineering, University of Virginia, Charlottesville, Virginia 22904, USA}

\author{George Kotsonis}
\affiliation{Department of Materials Science and Engineering, North Carolina State University, Raleigh, North Carolina 27695, USA}

\author{Ashutosh Giri}
\affiliation{Department of Mechanical and Aerospace Engineering, University of Virginia, Charlottesville, Virginia 22904, USA}

\author{Christopher T. Shelton}
\affiliation{Department of Materials Science and Engineering, North Carolina State University, Raleigh, North Carolina 27695, USA}

\author{Edward Sachet}
\affiliation{Department of Materials Science and Engineering, North Carolina State University, Raleigh, North Carolina 27695, USA}

\author{Zhe Cheng}
\affiliation{George W. Woodruff School of Mechanical Engineering, Georgia Institute of Technology, Atlanta, Georgia 30332, USA}

\author{Brian M. Foley}
\affiliation{George W. Woodruff School of Mechanical Engineering, Georgia Institute of Technology, Atlanta, Georgia 30332, USA}

\author{Zeyu Liu}
\affiliation{Department of Aerospace and Mechanical Engineering, University of Notre Dame, Notre Dame, Indiana 46556, USA}

\author{Shenghong Ju}
\affiliation{Department of Mechanical Engineering, The University of Tokyo, Bunkyo, Tokyo 113-8656, Japan}
\affiliation{Center for Materials research by Information Integration, National Institute for Materials Science, 1-2-1 Sengen, Tsukuba, Ibaraki 305-0047, Japan}

\author{Junichiro Shiomi}
\affiliation{Department of Mechanical Engineering, The University of Tokyo, Bunkyo, Tokyo 113-8656, Japan}
\affiliation{Center for Materials research by Information Integration, National Institute for Materials Science, 1-2-1 Sengen, Tsukuba, Ibaraki 305-0047, Japan}

\author{Mark S. Goorsky}
\affiliation{Department of Materials Science and Engineering, University of California, Los Angeles, California 90095, United States}

\author{Samuel Graham}
\affiliation{George W. Woodruff School of Mechanical Engineering, Georgia Institute of Technology, Atlanta, Georgia 30332, USA}
\affiliation{School of Materials Science and Engineering, Georgia Institute of Technology, Atlanta, Georgia 30332, USA}

\author{Tengfei Luo}
\affiliation{Department of Aerospace and Mechanical Engineering, University of Notre Dame, Notre Dame, Indiana 46556, USA}
\affiliation{Center for Sustainable Energy of Notre Dame (ND Energy), University of Notre Dame, Notre Dame, Indiana 46556, USA}

\author{Asegun Henry}
\affiliation{George W. Woodruff School of Mechanical Engineering, Georgia Institute of Technology, Atlanta, Georgia 30332, USA}
\affiliation{School of Materials Science and Engineering, Georgia Institute of Technology, Atlanta, Georgia 30332, USA}
\affiliation{Heat Lab, Georgia Institute of Technology, Atlanta, Georgia 30332, USA}

\author{Jon-Paul Maria}
\affiliation{Department of Materials Science and Engineering, North Carolina State University, Raleigh, North Carolina 27695, USA}

\author{Patrick E. Hopkins}
\email{phopkins@virginia.edu}
\affiliation{Department of Mechanical and Aerospace Engineering, University of Virginia, Charlottesville, Virginia 22904, USA}
\affiliation{Department of Materials Science and Engineering, University of Virginia, Charlottesville, Virginia 22904, USA}
\affiliation{Department of Physics, University of Virginia, Charlottesville, Virginia 22904, USA}

\date{\today}


\begin{abstract}
We present experimental measurements of the thermal boundary conductance (TBC) from $77 - 500$ K across isolated heteroepitaxially grown ZnO films on GaN substrates.  These data provide an assessment of the assumptions that drive the phonon gas model-based diffuse mismatch models (DMM) and atomistic Green's function (AGF) formalisms for predicting TBC.  Our measurements, when compared to previous experimental data, suggest that the TBC can be influenced by long wavelength, zone center modes in a material on one side of the interface as opposed to the ``vibrational mismatch'' concept assumed in the DMM; this disagreement is pronounced at high temperatures.  At room temperature, we measure the ZnO/GaN TBC as $490\lbrack +150, -110\rbrack$ MW m$^{-2}$ K$^{-1}$.  The disagreement among the DMM and AGF and the experimental data these elevated temperatures suggests a non-negligible contribution from additional modes contributing to TBC that not accounted for in the fundamental assumptions of these harmonic formalisms, such as inelastic scattering.  Given the high quality of these ZnO/GaN interface, these results provide an invaluable critical and quantitive assessment of the accuracy of assumptions in the current state of the art of computational approaches for predicting the phonon TBC across interfaces.
\end{abstract}

\maketitle

The thermophysical property defining thermal transport across an interface between two materials is often termed the thermal boundary conductance (TBC), frequently approximated through phonon gas theory as\cite{duda2010aa}
\eqn{
h_{\mt{K}}=\frac{1}{4}\sum_j\int\limits_{\textbf{k}} C_j\left(\textbf{k}\right)v_{g,j}\left(\textbf{k}\right)\zeta\left(\textbf{k}\right){{d}}\textbf{k}
}{eqn1}
where $j$ is the phonon polarization index, $\textbf{k}$ is the wave-vector, $v_{g}$ is the group velocity and $\zeta$ is the phonon transmission coefficient. According to Eq.~\ref{eqn1}, the maximum TBC across an interface can be achieved by engineering $\zeta$ to approach unity. However, the TBC is often assumed to decrease as the ratio of Debye temperatures of the materials comprising the interface decrease; in this case, the materials become more "vibrationally mismatched", and thus this transmission coefficient reduces.\cite{wilson2015aa,lyeo2006aa,stevens2005aa,costescu2003aa,stoner1992aa,cho2014aa,swartz1989aa} This vibrational matching concept is the basis of the traditionally assumed diffuse mismatch model (DMM),\cite{swartz1989aa} which, as with Eq.~\ref{eqn1}, is rooted in the assumptions of the ``phonon gas model'' (PGM).  Under this formalism, at a heterogeneous interface, $\zeta$ can never approach unity due to the differing vibrational densities of states.  However, recent work has demonstrated that at well-prepared epitaxial or well-bonded interfaces, the TBC can approach this maximal limit.\cite{costescu2003aa,wilson2015aa,giri2016aa} These experimental works draw into question the validity of the DMM, and other PGM-based approaches, in predicting the TBC across material interfaces, a question that has also been raised recently in computational studies.\cite{gordiz2016aa,gordiz2017aa}

In principle, a direct comparison of these models to experimental data should provide a check of the validity of these theoretical approaches to correctly predict the phonon driven TBC.  Indeed, several groups have recently provided this comparison to assess the validity of computational approaches based on the DMM or Atomic Green's Function (AGF) formalisms.\cite{ye2017aa,cheaito2015aa,dechaumphai2014aa} However, these works, along with the overwhelming majority of measurements of TBC across interfaces, have focused on metal/non-metal interfaces.\cite{hopkins2013aa,monachon2016aa,cheaito2015aa,stevens2005aa,stoner1993aa} Arguments rooted in the assumption that electron-phonon scattering at metal/non-metal interfaces will contribute to TBC have often been made to rectify models and data for metal/non-metal interfaces.\cite{majumdar2004aa,wang2012ab,lu2016ab,ordonez-miranda2011aa,sadasivam2015aa,huberman1994aa,sergeev1998aa,sergeev1999aa,sergeev2000aa,sergeev2002aa} While the validity of these electron-phonon assumptions have been questioned via selected experiments,\cite{lyeo2006aa,hohensee2015aa,giri2015ab,li2017ac} the presence of this potential interfacial heat transfer mechanism certainly calls into question the direct comparison of metal/non-metal TBC to the PGM or AGF models, effectively leaving these phonon TBC models un-vetted.  

Clearly, a measurement of TBC across an interface of two crystalline non-metals of high equality (e.g., epitaxially grown with little to no disorder/dislocations) would enable better assessment of the validity of the DMM and AGF for predicting phonon transport across interfaces.  However, previous works reporting measurements of this non-metal/non-metal TBC across single interfaces (i.e., not interpreted from superlattice measurements or across transition layer interfaces) are lacking and are limited to highly dislocated or amorphous interfaces.\cite{hopkins2010aa,hopkins2011aa,kimling2017aa,zhu2010aa} It is well known that interfacial disorder can lead to changes in TBC,\cite{hopkins2013aa} thus, these aforementioned non-metal/non-metal interface studies are not ideal to compare to phonon computational formalisms, which can not take into account the nonidealities at the interface.

In this work, we overcome this void in the literature by studying the TBC across ZnO/GaN interfaces.  We experimentally measure the TBC across isolated heteroepitaxially grown ZnO films on GaN substrates from $77 - 500$ K.  The relatively high lattice matching and epitaxial growth ensure high crystalline quality of the single crystalline ZnO near the GaN interface.  The measured TBCs are then compared directly to DMM and AGF simulations using first-principles-derived phonon dispersions.  Our measurements show relatively high values for the ZnO/GaN TBCs at elevated temperatures that exceed the values predicted by AGF and DMM calculations by nearly a factor of two at elevated temperatures.  This difference between experiment and computation suggests the basic assumptions rooted in these formalisms are not suitable to predict the phonon TBC; this points to the potential existence of anharmonic phonon interactions enhancing the TBC at this ZnO/GaN interface, a process that is not rigorously accounted for in DMM or AGF simulations. We compare our measured ZnO/GaN TBC to calculations of the theoretical maximum predicted under various assumptions, along with previously derived TBCs from measurements of ZnO/hydroquinone (HQ) superlattices.\cite{giri2016aa} In comparison to the various models and previous data, our results suggest that the TBC can be intrinsic to the phonon modes in the ZnO, and not necessarily related to a ``transmission'' of modes restricted by the vibrational states on the other side of the interface.  This nanoscopic mechanism of TBC can not be predicted by PGM-based formalisms.  

\begin{figure}
\begin{center}
\includegraphics[width=3.3in]{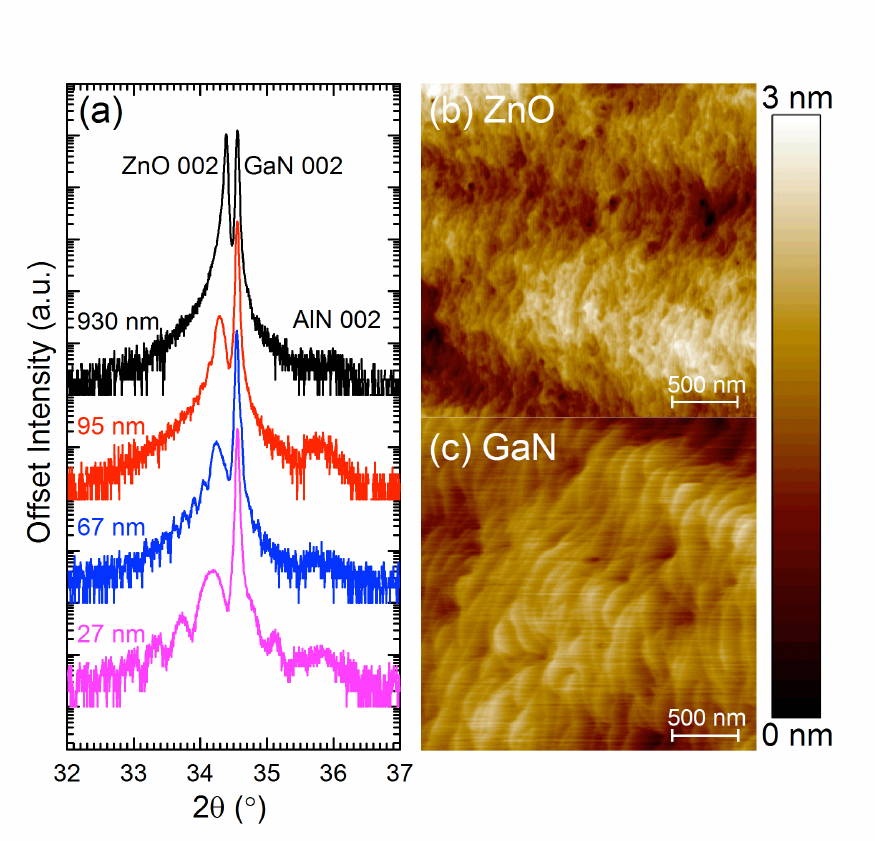}
\caption{(a) XRD patterns for ZnO grown on GaN with various thicknesses, (b) AFM data for 95 nm of ZnO, and (c) AFM data for the bare GaN wafer.}
\label{Characterization}
\end{center}
\end{figure}

ZnO thin films of thicknesses from $5 - 930$ nm were grown heteroepitaxially on a Ga-polar GaN wafer by pulsed-laser deposition.  The GaN wafer was prepared on a $[0001]$-sapphire wafer by metal-organic chemical vapor deposition employing an AlN buffer layer.\cite{collazo2006aa, mita2008aa}  Film structure, roughness, and thickness were characterized by X-ray diffraction (XRD), atomic force microscopy (AFM), and X-ray reflectivity (XRR) respectively.  Figure 1a shows offset $2\theta-\omega$ XRD scans of the ZnO and GaN $002$ reflections for a representative subset of the films.  A strong ZnO 002 reflection can be seen beside the GaN $002$ peak, indicating the films adopt the $[001]$ orientation of the GaN.  Pendellosung fringing can be seen in some of the XRD patterns, a result of X-ray interference from the ZnO thickness and usually indicative of a high crystal quality and smooth interfaces.\cite{Kato1959aa}  A $2\theta$-peak shift to lower angles (larger interplanar spacing at lower thickness) can also be seen in Fig.~\ref{Characterization}a.  This shift implies that thin ZnO films experience in-plane compressive epitaxial strain, while thicker ZnO films tend to relax toward its bulk lattice parameter.  Long range $2\theta-\omega$ scans (see Supplementary Information) indicate a small amount of $[110]$-oriented ZnO grains at thicknesses above $100$ nm, but thinner films exhibit a pure $[001]$ orientation, implying that the $[110]$ ZnO nuclei precipitated away from the ZnO/GaN interface.  In this regard, we assume the ZnO/GaN interface is nearly identical and epitaxial for each of the films in this study.

Representative AFM data for the 95 nm thick ZnO film and the GaN wafer are shown in Figs.~1b and 1c, respectively.  The GaN surface exhibits a step-terrace morphology and the ZnO films adopt a comparable morphology with less distinct, but still observable, step edges.  The RMS surface roughness of all ZnO films were $\sim1$ nm or less, and that of the GaN was $<1$ nm, determined by AFM image analysis.  Together, the XRD and AFM data suggest heteroepitaxy and smooth and coherent ZnO/GaN interfaces.  For ZnO films less than 100 nm thick, XRR was employed for thickness determination.\cite{Wainfan1960aa} Fitting of the XRR data provided ZnO layer thickness and yielded surface roughness that agree with AFM data.  For the $180$ and $930$ nm thick films, XRR was not able to resolve thickness oscillations and, as such, selective etching and AFM profilometry of a companion film, grown in the same growth as the thermally characterized films, were used to determine the thickness.

Our thermal measurements are carried out using time domain thermoreflectance (TDTR), a technique that is described in detail elsewhere.\cite{cahill2004aa,schmidt2008aa,hopkins2010aa} We first measure a piece of the same GaN on sapphire wafer on which the ZnO films were grown in order to measure both the GaN/sapphire TBC as well as the thermal conductivity, $\kappa$, of the GaN.  We measure $\kappa_{\mt{GaN}}=159\pm$12 W m$^{-1}$ K$^{-1}$, in line with previous measurements of high quality GaN with thickness of $\sim 1\, \mu$m.\cite{beechem2016aa}  These values are used in all subsequent analyses discussed below and has the effect of reducing the number of unknowns in our analysis to: $h_\mt{K,Al/ZnO}$, $\kappa_\mt{ZnO}$, and $h_\mt{K,ZnO/GaN}$.  In order to determine the thermal conductivity of our ZnO, we measure a $930$ nm thick film.  This thickness ensures we are only sensitive to the thermal conductivity of the ZnO and $h_\mt{K,Al/ZnO}$, and not to $h_\mt{K,ZnO/GaN}$.  The thick ZnO film yields a thermal conductivity of $\kappa_\mt{ZnO} =53.4\pm$4 W m$^{-1}$ K$^{-1}$, similar to values found in literature for high quality ZnO (Refs.~\onlinecite{alvarez2010aa,tsubota1997aa,barrado2004aa}) and in line with recent computational work.\cite{wu2016aa}  Both the GaN control and thick ZnO thermal conductivity were independently tested and verified via TDTR at the University of Virginia and Georgia Institute of Technology.  It should be noted there are a host of lower literature values for thin film ZnO that are highly influenced by the microstructural features present in the films, namely the presence of grain boundaries, which may act as thermal scattering sites.\cite{huang2011aa,xu2012aa} 

\begin{figure}
\begin{center}
\includegraphics[width=3.3in]{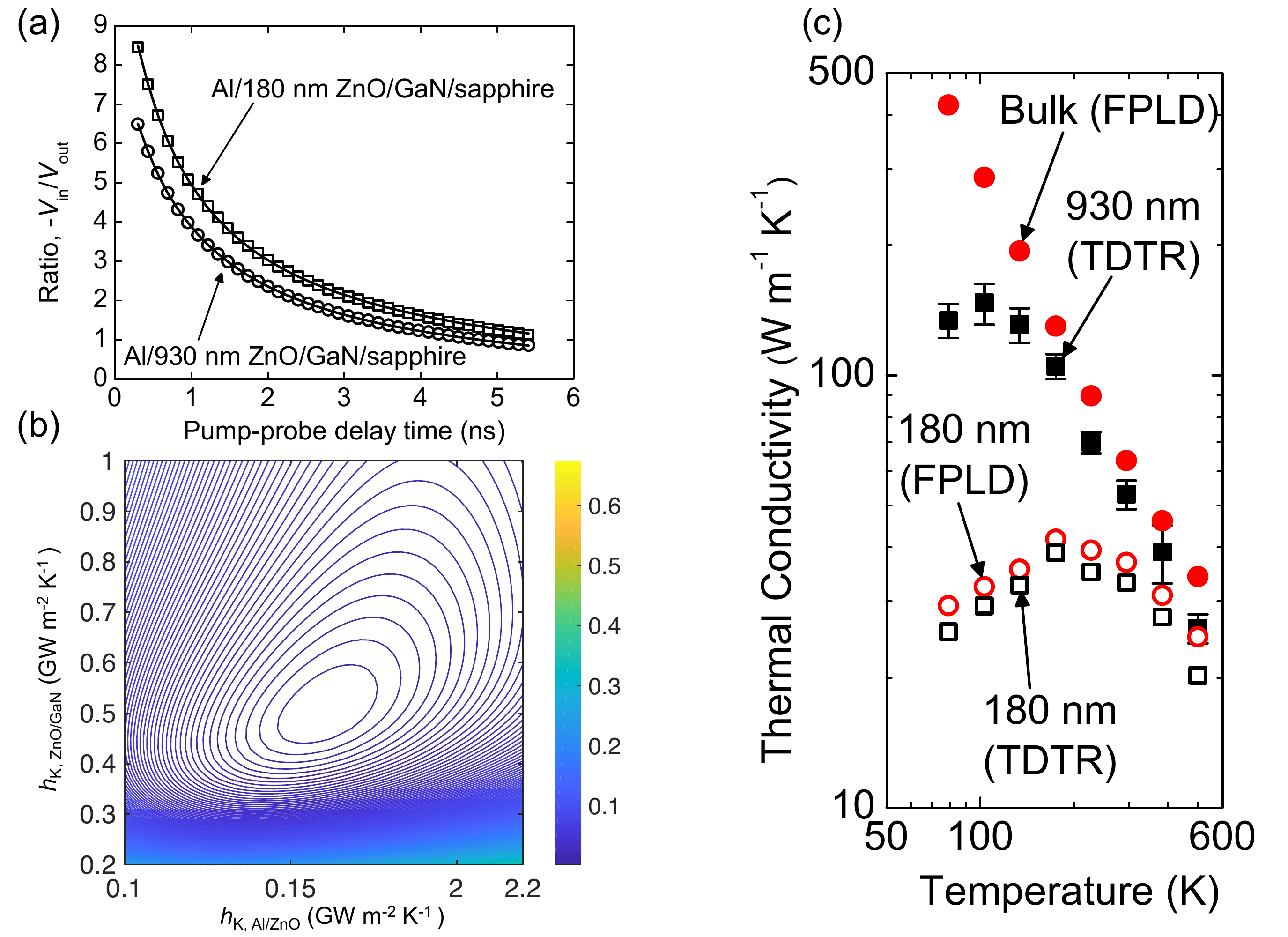}
\caption{(a) Experimental data (open symbols) and the best fits to our thermal model (solid lines) for the 180 and 930 nm thick ZnO films. (b) Sensitivity contour plots for the 180 nm thick film for $h_\mt{K,ZnO/GaN}$ as a function of $h_\mt{K,Al/ZnO}$. The lowest value of the contour lines indicates the combinations of thermal parameters that lie within a 95\% confidence interval.  As is clear from the sensitivity contour plot, a relatively confined range of values for $h_\mt{K,ZnO/GaN}$ and $h_\mt{K,Al/ZnO}$ can produce best fits to the TDTR data, confirming our uncertainty bounds in our measurements of $h_\mt{K,ZnO/GaN}$ in the $180$ and $95$ nm thick films. (c) Measured thermal conductivity of the 930 nm ZnO films (filled squares) compared to predictions of bulk ZnO thermal conductivity via first principles lattice dynamics (FPLD, filled circles).  Using a series resistance model that accounts for $h_{\mt{K,ZnO/GaN}}$, we then compare the effective thermal conductivities of the measured 180 nm films (open squares) and that predicted from FPLD (open circles).  The agreement supports our analysis procedure that assumes the thermal conductivity of the ZnO is reduced due to a TBC at the ZnO/GaN interface.\cite{zeng2001aa}}
\label{Fig2}
\end{center}
\end{figure}

In order to obtain $h_\mt{K,ZnO/GaN}$, we test films with thickness of $180$ and $95$ nanometers.  Taking $\kappa_\mt{ZnO}$ from the thick film leaves us with two unknown parameters in the thermal model, $h_\mt{K,ZnO/GaN}$ and $h_\mt{K,Al/ZnO}$, which we can determine by fitting $h_\mt{K,Al/ZnO}$ with the in-phase signal, $V_{\mt{in}}$ signal and $h_\mt{K,ZnO/GaN}$ with the ratio of $-V_{\mt{in}}/V_{\mt{out}}$.  We iterate these values into the opposing thermal models until the values converge, yielding an average for these films of moderate thickness of $h_\mt{K,ZnO/GaN} =$ $490\lbrack +150, -110\rbrack$ MW m$^{-2}$ K$^{-1}$. Exemplary TDTR data and the model fits for the $930$ and $180$ nm films are shown in Fig.~\ref{Fig2}a.  Figure \ref{Fig2}b shows the results from a contour plot analysis that demonstrates the mean square deviation of the thermal model to the TDTR data ($-V_{\mt{in}}/V_{\mt{out}}$) for various combinations of $h_\mt{K,ZnO/GaN}$ and $h_\mt{K,Al/ZnO}$  as input parameters in the model for the 180 nm film.  The lowest value of the contour lines indicates the combinations of thermal parameters that lie within a 95\% confidence interval.  As is clear from the sensitivity contour plot, a relatively confined range of values for $h_\mt{K,ZnO/GaN}$ and $h_\mt{K,Al/ZnO}$ can produce best fits to the TDTR data, confirming our uncertainty bounds in our measurements of $h_\mt{K,ZnO/GaN}$ in the $180$ and $95$ nm thick films. We further confirm these reported values for $h_\mt{K,ZnO/GaN}$ through measurements on the thinner ZnO films with thicknesses of $5$, $10$, $19$, $27$, $42$, and $66$ nm, discussed in the Supplementary Information.

To allow for further investigation of the transport properties at this well matched interface, we turn to temperature dependent measurements of $h_\mt{K,ZnO/GaN}$.  Relevant thermophysical properties for the temperature dependent measurements and analysis were taken from a combination of measurements on control samples (thick GaN and thick ZnO) and a variety of existing literature.\cite{dobrovinskaya2009aa, ditmars1982aa, kremer2005aa, simon2014aa, alvarez2010aa, madelung1999aa, corruccini1960aa}  As a validation of our analysis procedure, we compare our measured thermal conductivities of $\kappa_{\mt{ZnO}}$ to those predicted via first principles lattice dynamics (FPLD).\cite{wu2016aa} As shown in Fig.~\ref{Fig2}c, the measured thermal conductivities of our thickest ZnO film ($930$ nm) agrees well with the FPLD predictions at higher temperatures.  At cryogenic temperatures, the deviation between the TDTR data and FPLD predictions is most likely due to size effects in the $930$ nm that are more pronounced at these lower temperatures.  To test our assumptions regarding the role of size effects and our ability to extract the ZnO/GaN TBC in our TDTR analysis, we calculate $\kappa_{\mt{ZnO}}$ for the $180$ nm film using FPLD with a boundary resistance in series, which we take from our measured data, given by $\kappa_{\mt{effective}}=\left(180\times 10^{-9}\,\mt{m}/\kappa_{\mt{FPLD,bulk}}+1/h_\mt{K,ZnO/GaN}\right)^{-1}$.  We compare these predictions to the measured effective thermal conductivities via TDTR.  The agreement between these FPLD predictions and our measured data shown in Fig.~\ref{Fig2}c give additional credence to our TDTR fitting procedure used to measure $h_\mt{K,ZnO/GaN}$ over a range of temperatures; namely that the thermal conductivity of the ZnO thin films is reduced due to the finite TBC at the ZnO/GaN interface.\cite{zeng2001aa} We ascribe the slight disagreement between the FPLD model and our TDTR data to atomic defects in the ZnO, where the FPLD simulations are based on a perfect crystal.

\begin{figure}
\begin{center}
\includegraphics[width=2.75in]{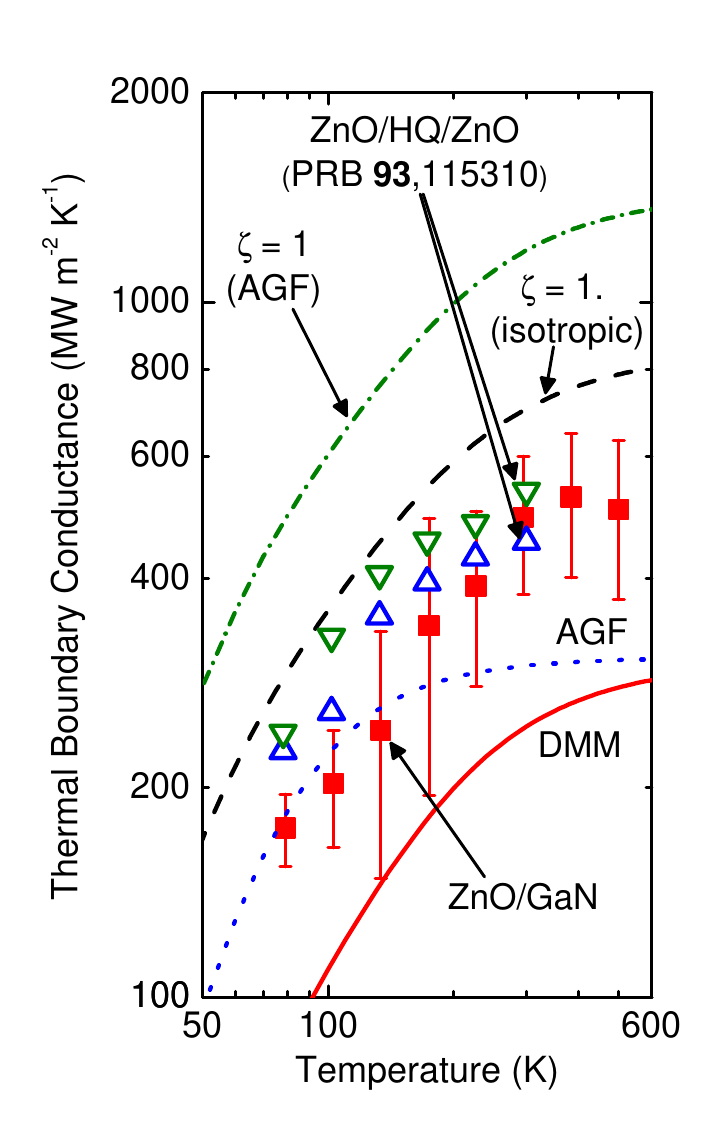}
\caption{Measured ZnO/GaN TBC as a function of temperature (filled squares) compared to previous data measured across ZnO/HQ/ZnO interfaces (open triangles),\cite{giri2016aa} DMM and AGF predictions of the ZnO/GaN TBC (solid and dotted lines, respectively), and the maximum TBC predicted via Eq.~\ref{eqn1} (assuming isotropy, dashed line) and this maximum TBC assuming a more precise shape to the Brillouin Zone (dot-dashed line).\cite{wu2016aa} At low temperatures, the agreement between the data and AGF, and their disagreement with the DMM is ascribed to the failure of the DMM to account for the effective interfacial transport of long wavelength modes. The disagreement between the measured data and both the AGF and DMM (and the convergence of these models) suggest that inelastic scattering among different mode energies could be contributing to the ZnO/GaN TBC, a phenomenon that is not rigorously accounted for in our DMM and AGF formalisms assumed here.}  
\label{TempResults}
\end{center}
\end{figure}

Figure \ref{TempResults} shows the results for $h_\mt{K,ZnO/GaN}$ as a function of temperature.  We also plot experimental data of TBC across ZnO/HQ/ZnO interfaces extracted from thermal conductivity measurements of organic/inorganic multilayers grown from atomic/molecular layer deposition from our recent work.\cite{giri2016aa} The similarities in TBCs, in both magnitude and temperature trends, suggest similar interfacial heat transport mechanisms driving the TBC.  In the case of the ZnO/HQ/ZnO, the results suggested that the TBC was driven by the phonon flux in the ZnO. The similarity in our current values for this heterogeneous ZnO/GaN interface suggest the same: namely, the heat transport mechanisms driving the TBC across this ZnO/GaN epitaxial interface are intrinsic to the ZnO, an observation that is contradictory to the DMM and other PGM-based theories.  Recent works\cite{duda2011aa,hua2017aa} have suggested that the TBC across ideal interfaces can be driven by near perfect transmission of long wavelength, zone center phonons. This is in disagreement to theories rooted in the PGM that assume vibrational mismatch between two materials' densities of states can impact transmission of all phonon wavelengths, such as the DMM and other Landauer-based TBC that do not capture the wave-based nature of phonon transport.\cite{imry1999aa}  Our results support these aforementioned theories,\cite{duda2011aa,hua2017aa} and suggest that at ``perfect'' interfaces, the conductance can be intrinsic to one of the materials adjacent to the interface, in our case, the ZnO.

We more quantitatively analyze our measured TBC data and study this potential failure of AGF and PGM-based DMM assumptions by calculating the TBC via the DMM and AGF. First, we calculate the ZnO/GaN TBC with the DMM assuming an isotropic Brillouin Zone, an assumption that is arguably the most widely applied in DMM predictions.\cite{swartz1989aa,costescu2003aa,duda2010aa,duda2010ab} The details of these calculations are further discussed in the Supplementary Information, but we note that we calculate the DMM via a polynomial fit to the phonon dispersion of ZnO and GaN in the $\Gamma\rightarrow M$ direction.\cite{ruf2001aa,serrano2010aa}  The DMM underpredicts the measured $h_\mt{K,ZnO/GaN}$ by nearly a factor of two across the entire temperature range.  It should be noted that our assumption of Brillouin Zone isotropy may certainly be playing a role in this disagreement, as anisotropy in the crystal structure can affect TBC.\cite{duda2009aa,duda2010ac,hopkins2011ac,chen2013aa} However, as previously discussed, our data also suggest that the fundamental assumptions driving the DMM can not capture the TBC at this epitaxial ZnO/GaN interface, so this disagreement is not surprising.

We also calculate the ZnO/GaN TBC using AGF, as shown in Fig.~\ref{TempResults}. Our AGF calculations include the exact atomic level detail of the interface, in comparison to the DMM which is limited in this atomic-level description. We note that direct comparisons between AGF calculations and an appropriately matched experimental measurement of an isolated nonmetal-nonmetal TBC are, to the best of our knowledge, nonexistent.  Thus, our results herein for an AGF prediction of the ZnO/GaN interface provide a critical comparison that has been absent in the literature.  Our AGF calculations were performed \textit{ab initio}, using density function theory (DFT). The electronic structure calculations were performed using the Vienna Ab Initio Software Package (VASP).\cite{kresse1994aa,kresse1993aa}  The general details associated with the AGF implementation are well described elsewhere,\cite{wang2006aa, zhang2007aa, tian2012aa, li2012aa} and our specific assumptions are outlined in the Supplementary Information; the results of these calculations are shown in Fig.~\ref{TempResults}.  In general, the AGF calculations capture the low temperature values but underpredict the trends and values at high temperatures.  

Unlike the DMM, the AGF formalism accounts for the wave-like nature of phonon transport and naturally captures phonon transport processes typically associated with traditionally assumed acoustic mismatch theories;\cite{little1959aa,snyder1970aa,swartz1989aa} that is, AGF can account for the fact that long wavelength phonons can efficiently transfer energy across interfaces more so than short wavelength phonons,\cite{hopkins2009ad,latour2017aa} a phenomena that has been theorized previously.\cite{hua2017aa,hopkins2011ai,duda2012ab,hopkins2011aa} It is of note that the DMM does not account for this effect and assumes all phonons scatter diffusively at interfaces, thus underpredicting the contribution of long wavelength phonons to TBC.  This can potentially explain the disagreement between the AGF and DMM predictions at low temperatures, and supports these aforementioned previous theories that long wavelength phonons can effectively transfer energy across heterogeneous interfaces,\cite{hua2017aa,hopkins2011ai,duda2012ab,hopkins2011aa} and do not obey DMM-based constraints.

At higher temperatures the DMM and AGF calculations converge, while still underpredicting the experimental data by nearly a factor of two.  A potential source of this underpredicition has often been ascribed to inelastic scattering at the interface, where anharmonic interactions among multiple phonons can open up additional parallel pathways for increases to TBC.\cite{duda2011aa,hopkins2009ac,hopkins2011ab,lyeo2006aa,hopkins2008ae,panzer2010aa,saaskilahti2014aa}  Given that both our AGF and DMM calculations only assume harmonic interactions, this indeed could explain the discrepancy between the models and our measured data.  More specifically, especially given the rigor of our AGF calculations and the quality of this ZnO/GaN interface, inelastic scattering processes are most likely contributing to the TBC at the ZnO/GaN interface at elevated temperatures. 

Figure \ref{TempResults} also shows the calculations for the maximum possible TBC, calculated both via Eq.~\ref{eqn1} with $\zeta$=1, which assumes an isotropic Brillouin Zone in the $\Gamma\rightarrow M$ direction, and using AGF at a ZnO/ZnO interface, which accounts for the exact geometry of the Brillouin Zone in ZnO.  The most accurate calculation for this maximum TBC determined via the AGF calculations sets the upper bound for the ZnO flux, and our data are over a factor of two lower than this limit.   We note the substantial disagreement between the AGF maximum limit and that calculated via Eq.~\ref{eqn1}, which is most likely due to our assumptions of the Brillouin Zone shape and the dispersion relation used to calculate TBC under the DMM framework, as previously discussed.  Thus, when determining the maximum possible TBC across interfaces, it is important to use as detailed a phononic spectra as possible to ensure accuracy.  Furthermore, even at these near perfect epitaxial ZnO/GaN interfaces, the measured TBC is only $\sim 30 \%$ of the maximum TBC. Given our previous discussion regarding the high efficacy of long wavelength modes, these data also suggest that high frequency modes are not effective carriers of energy across interfaces, which supports recent computational findings.\cite{latour2017aa,hua2017aa}

In summary, we have reported on experimental measurements of the TBC across isolated heteroepitaxially grown ZnO films on GaN substrates from $77 - 500$ K, providing a direct comparison of measured TBCs to DMM and AGF simulations.  This comparison allows for a direct assessment of the assumptions implemented in DMM and AGF calculations.  Our measurements, when compared to previous experimental data across ZnO/HQ/ZnO interfaces, suggest that the TBC can be influenced by the modes in the material on one side as opposed to the ``vibrational mismatch'' concept assumed in the DMM.  Furthermore, the disagreement between both the DMM and AGF and the experimental data at elevated temperatures suggest the contribution of additional modes of TBC not accounted for in the fundamental assumptions of these harmonic formalisms, such as inelastic scattering.  Our data suggest that long wavelength phonons can effectively transmit across interfaces, supporting the findings of recent computational studies.\cite{latour2017aa,hua2017aa}

These results also provide invaluable impact into strategies of thermal mitigation and power dissipation electronic devices such as microprocessors, semiconductor-based radio frequency devices, and radar amplifiers.\cite{smith2004aa,trew2002aa,lin2008aa} This thermal bottleneck in these devices has proven to be the major roadblock in achieving higher power gallium nitride high-electron mobility transistors (GaN HEMTs) along with other semiconductor-based high-frequency, high-output power technologies.\cite{mishra2002aa,zhou2013aa} A significant limitation in the ability to scale devices to higher powers, especially as active regions continue to dimensionally shrink, is the TBC at GaN interfaces, and its effect on heat transfer into sub-mounts and heat sinks.\cite{cho2013aa}  Our results lend insight into how phonon modes couple energy across GaN-based interfaces.  

We appreciate support from the Office of Naval Research, Grant No.~N00014-15-12769. 

\bibliography{ZnO_GaN_BIB}

\end{document}